\documentclass[10pt]{article}
\usepackage{geometry}
\usepackage{cite}
\usepackage{amssymb,amsmath}
\usepackage{wrapfig}
\usepackage{graphicx}
\usepackage[small]{caption}
\usepackage[titletoc,title]{appendix}
\usepackage{authblk}
\usepackage{abstract}

\begin{document} 
\title{On possible extensions of quantum mechanics}
\author{Yiruo Lin \thanks{yiruolin@tamu.edu, linyiruo1@gmail.com}}
\affil{\textit{Department of Computer Science \& Engineering, Texas A\&M University}}
\date{}
\maketitle

\begin{abstract}
\normalsize
It was argued \cite{free choice ncomm} that there can be no extension of quantum mechanics with improved predictive power on a measurement freely chosen, independently of any event that is not in its future light cone. The assumption of measurement choice was criticized \cite{criticism} to be too strong to be physically necessary and extensions of quantum mechanics were shown \cite{extension} to be possible under a more relaxed measurement assumption. Here I point out an error in the criticism and observe that the actual mistake of the no-go theorem lies in an unwarranted assumption implicitly made in the proof of \cite{free choice ncomm}. Hence, quantum mechanics is guaranteed to have the maximal predictive power only in situations of complete certainty and complete uncertainty about measurement outcomes. I then show that the measurement assumption can be further relaxed without affecting the conclusion on the predictive power of quantum mechanics versus alternative theories. I further study the optimal predicative improvement over quantum mechanics of local spin measurements on a pair of entangled qubits by any alternative theory and conjecture a strict upper bound.  
\end{abstract}

\section{Introduction}

The completeness of quantum mechanics has been a long-standing problem, due to the nondeterministic randomness feature of quantum mechanical predictions on general measurement outcomes. In \cite{free choice ncomm}, Colbeck and Renner (\textbf{CR}) argued that no extension of quantum mechanics is possible in the sense that no alternative theory can provide more accurate prediction on any measurement. The argument is based on a free measurement choice assumption (\textbf{FR assumption}) that a measurement setting can be chosen independently of any event lying outside its future light cone. Unsatisfied with the FR assumption, Ghirardi and Romano(\textbf{GR}) \cite{criticism} argued that a physically more sensible assumption of measurement settings (\textbf{FW assumption}) only requires measurements be chosen 1) independently of each other for spacelike separated observers and 2) independently of any hidden variable (or ontic state) that specifies the physical state of the system under measurements. Under FW assumption together with the usual non-signaling conditions, GR constructed ontological models (models using the putative ontic states to give more complete descriptions of microscopic world than quantum mechanics) \cite{extension} which give explicitly different predictions from quantum mechanics while stay consistent with quantum mechanical statistics when integrating out hidden variables (ontic states). GR attributed the apparent conflict with the no-go theorem by CR to the adoption of different assumptions (FR vs. FW assumption). \\

In this paper, I will show that the difference in the conclusions by CR and GR lies not in the assumptions they made, but due to an implicit assumption by CR which is flawed. In fact, the conclusion on the extensibility of quantum mechanics is independent of FR vs. FW assumption. Without affecting the conclusion, the assumptions can be relaxed further, requiring only non-signaling between spacelike  observers. Under the most general assumptions, quantum mechanics is optimal in predicting experiment outcomes only in situations of complete measurement certainty and complete measurement uncertainty. While the former case is trivial as the prediction from quantum mechanics is already optimal and deterministic, the latter case corresponds to uniform distributed probabilities over measurement outcomes. Away from these two special cases, improvement over predictions of quantum mechanics is possible. As shown by GR \cite{extension}, quantum mechanical correlations set constraints on the degree of improvement by a generic theory. Their analysis can be readily generalized to more general situations without changing the constraints. I will then consider local spin measurements on a qubit entangled with another a qubit and study the maximal possible improvement in predicting the measurement outcomes of 1) any alternative theory and 2) a class of ontological models. Finally, I conjecture a strict upper bound on the improvement.     \\

The paper is organized as follows. I will first recapitulate the derivation of the no-go theorem by CR in section \ref{CR theorem}, followed by the criticism and the counterexamples to the no-go theorem by GR in section \ref{criticism}. In section \ref{observation}, I will then pinpoint the essential dispute between the two groups of authors and observe the actual mistake made by an implicit assumption in the derivation of the no-go theorem at the stage of generalizing situations of complete uncertain measurement outcomes to general measurement outcome probabilities. In section\ref{generalization}, I will generalize the analysis in \cite{extension} of the extensibility of quantum mechanics to the most general types of ontological models and to the more relaxed constraints on the spacelike observers. In section \ref{improvement}, I analyze in general the optimal improvement in predicting the spin measurement outcomes on a qubit from an entangled qubit pair, given the constraints set by quantum mechanics. I also obtain the optimal performance of a class of ontological models in predicting such measurements. Finally in section \ref{Conclusions}, the main findings will be summarized with further comments including a conjecture on a strict upper bound of prediction improvement over quantum mechanics.

\section{the No-go theorem by CR} \label{CR theorem} 

The problem addressed by CR is the following: Is it ever possible to make more precise predictions on outcomes $X$ of measurements $A$ preformed on a microscopic system $S$ than quantum mechanics when supplemented with additional information $\Xi$ beyond the (pure) quantum state $\psi$ of the system? The improvement over quantum mechanics can be quite general and it needs not complete quantum mechanics so as to give completely deterministic predictions. The authors related outcome statistics of the local measurements $A$ on the system $S$ to correlations between outcomes of local measurements $A$ and $B$, which are spacelike separated from each other. This can be achieved by assuming that a measurement $A$ on $S$ produces entanglement between $S$ and the measurement device $D$, according to  quantum mechanics. Further assume the entanglement between $S$ and $D$  can be preserved while being spatially separated so that spacelike separated measurements $A$ and $B$ on $S$ and $D$ can be performed respectively. \\

CR then made the FR assumption that each measurement choice of $A$ and $B$ can be chosen independently of any event not in their respective future light cones, namely, $P_{A|BCYZ}=P_A$ (similarly for $B$ and $C$), where $Y$ denotes the outcome of measurement $B$, $C$ and $Z$ correspond to extra measurement and outcome from which observers can retrieve supplemented information encoded in $\Xi$. Under the FR assumption and the non-signaling conditions ($P_{YZ|ABC}=P_{YZ|BC}$ for $A$ and similarly for $B$ and $C$) derived from it \cite{consistent}, CR proved that no information in $\Xi$ can provide improved prediction on outcome $X$ of measurement $A$ on system $S$ for effectively local spin measurements on subsystems $S$ and $D$ forming a pair of maximally entangled qubits. This is done by considering a set of local spin measurements $A$ and $B$ yielding joint probabilities $P_{XY|AB}$ whose values are prescribed by quantum mechanics. CR were able to show that $P_{X|Az}=P_{X|A}$ for any $z$  (lower case z denotes all possible values taken by $Z$), making use of a correlation measure of $P_{XY|AB}$ whose value vanishes for a pair of maximally entangled qubits. Hence, no extra information $\Xi$ can provide measurement outcome statistics different from quantum mechanics (i.e., the uniform distribution $P_{X|A}=1/2$, $X=\pm1$ corresponding to projection onto opposite spin orientations) when the measurements are on the local spin constituents of a pair of maximally entangled qubits.
 \\

To generalize the above conclusion to non-uniform distributions of measurement outcome statistics $P_{X=x_j|A}=P_j$ ($\sum_jP_j=1$ and $\exists j, j'$, $P_j\neq P_{j'}$) with the corresponding entangled state  $\sum_j\sqrt{P_j}|S_j\rangle|D_j\rangle$, CR appended extra measurement outcomes $X'$ such that the pair of outcomes $(X,X')$ are uniformly distributed. The new system forms multiple pairs of maximally entangled qubits on which transformed measurements are performed, yielding $(X,X')$. Since no extra information can give different measurement statistics from quantum mechanics for maximally entangled qubit pairs, CR concluded that the same conclusion applies to general situations of non-uniform distributions. \\


\section{Criticism by GR and counterexamples to the no-go theorem} \label{criticism}

While agreeing with the derivations of CR, GR questioned the physical necessity of the FR assumption made by CR. In light of the fact that the no superluminal signaling is the essential constraint preventing a fully deterministic description of a local system entangled with another local system spacelike separated from it, GR pointed out that the key to the completion of quantum mechanics lies in the accessibility of hidden variables (ontic states), instead of the free measurement choices required by the FR assumption. Indeed, the FR assumption rules out any deterministic completion of quantum mechanics trivially due to the non-signaling conditions. Hence, the proof of the CR no-go theorem is relevant for non-deterministic improvement over quantum mechanics. GR stressed that a deterministic completion of quantum mechanics is possible in principle such as the Bohmian mechanics, but is impossible in practice (experimentally) thanks to the non-signaling conditions. The hidden variables (e.g., the particle position) in the Bohmian mechanics are completely inaccessible to an observer so that no information of the hidden variables is available for superluminal signaling and no better prediction over quantum mechanics can be made by the observer. \\

GR discussed the implication of the FR assumption on free will of observers to choose experimental settings. They argued that violation of the FR assumption doesn't imply the lack of freedom of the observers to choose their experiments at will. To replace the FR assumption, they then put forth the FW assumption which only requires the mutual independence of measurement choices between spacelike observers and all possible underlying ontic states $\Lambda$ (of which the accessible part is given by $Z$), i.e., $P_{AB\Lambda}=P_AP_BP_\Lambda$. GR then decomposed the FR assumption into three parts: 1) the FW assumption, 2) the usual non-signaling conditions $P_{X|AB}=P_{X|A}$ (and similarly for $B$) and 3) the static condition (\textbf{ST condition}) of extra information $Z$: $P_{CZ|ABXY}=P_{CZ}$. GR attributed the no-go theorem of CR to assumption 3) which is unnecessarily strong. \\

In a subsequent work \cite{extension}, GR constructed a class of deterministic ontological models satisfying the first two parts of the FR assumption, but not the ST condition. In order to comply with the no superluminal signaling conditions, not all information of the underlying ontic states can be accessed. Following \cite{Leggett}, GR separated the ontic states $\Lambda$ into accessible and inaccessible parts. After integrating out the inaccessible part of $\Lambda$, the accessible part obeys the non-signaling conditions. The resulting model is probabilistic and the accessible part of $\Lambda$ yields measurement statistics which stays the same as quantum mechanics for a pair of maximally entangled qubits, but deviates from quantum mechanics for non-maximally entangled states. \\

\section{What's wrong with the CR no-go theorem?} \label{observation}

It appears that the central disagreement between CR and GR lies in the assumption concerning the correlation between $Z$ and $ABXY$ as implied by the ST condition. However, it turns out that the decomposition by GR of the FR assumption is incorrect. As already note by CR\cite{reply}, the ST condition is never assumed in their derivation of the no-go theorem. The decomposition is sufficient to derive the FR assumption but not necessary. Furthermore, the ST condition is trivially violated if the supplemented information $Z$ were to yield measurement statistics different from quantum mechanics. This can be readily seen by noting that the ST condition implies that $P_{XY|ABCZ}=P_{XY|AB}$. \\

In fact, one can readily check that the FR assumption is actually satisfied for the GR ontological models \cite{extension} after integrating out the inaccessible part of $\Lambda$. So why the two groups of authors reach contradictory results from essentially the same assumption? The answer lies in an implicit assumption made by CR at the stage of extending the no-go theorem from maximally entangled states to generally entangled states. To extend the scope of the no-go result, one has to transform general local measurements on a generally entangle state to new local measurements that probe a transformed system composed of maximally entangled pairs of qubits. Surely, for the transformed system, the no-go theorem again applies. However, one has to assume that in the process of the transformation, the predicability of any alternative theory on measurement outcomes changes in the same way as quantum mechanics. Logically, one has to assume what one wants to prove. Hence, the assumption is unwarranted and the generalization of the no-go result is invalid. \\

Indeed as already demonstrated by the ontological models of GR, the rate of change of predicative power of an alternative theory on local measurement outcomes can differ from quantum mechanics as the degree of entanglement of the system of interest changes (see Fig. 1 of \cite{extension} for a quantitative illustration). When the system is maximally entangled, any alternative theory yields the same predicability on any local measurement, i.e., complete uncertainty as predicted by quantum mechanics. As the system becomes less entangled, an alternative theory can give more information than quantum mechanics, i.e., yields different outcome statistics of a local measurement from quantum mechanics while complying with it after averaging over the ontic states. When the local system is an eigenstate of the local measurement, the whole system is in a product state of the local system and the experimental apparatus, and is completely unentangled. In this case, no alternative theory can give better prediction for the trivial reason. \\

To grasp an intuitive picture of the performance between an alternative theory and quantum mechanics, it may help to imagine a process of local spin measurements on a qubit in a definite spin orientation. When the measurement is projecting on directions orthogonal to the spin, the outcome is completely uncertain. After the measurement, the qubit becomes maximally entangled with the measurement device, forming effectively a pair of maximally entangled qubits. In this case, the CR no-go theorem applies and no other theory can give better predictions. Now perform a series of spin measurements by gradually rotating the orientation till aligning with the orientation of the qubit, the outcomes become more and more predictable and the qubit becomes less and less entangled with the measurement device. In the process, the predicability of an alternative theory can improve more than quantum mechanics and eventually converges to it. \\


\section{General constraints set by quantum mechanics on any alternative theory} \label{generalization}

I now generalize the work of GR\cite{extension} to show that the general constraints set by quantum mechanics stay the same when 1) the FW assumption by GR is removed and only the non-signaling conditions are required, 2) allowing an alternative theory to be nondeterministic at the deepest level.  \\

Consider a bipartite system with local measurements $A(a)$ and $B(b)$ performed on each subsystem. The lower case $a$ and $b$ refer to local measurement settings and the upper case $A$ and $B$ denote the measurement outcomes. Suppose the bipartite system is in a pure quantum state $\psi$ which is associated with a statistical distribution $\rho_{\psi}(\lambda)\geq0$ of the underlying ontic states $\lambda$. The predictions given by the ontic description should be consistent with quantum mechanics. For local average of $A$, 
\begin{eqnarray}
\int A_{\psi}(a,b,\lambda)\rho_{\psi}(a,b,\lambda)d\lambda=\langle A(a)\rangle_\psi, \label{A_ave}
\end{eqnarray}
and similarly for B. $A_{\psi}(a,b,\lambda)$ is the outcome given by ontic state $\lambda$, which in general could depend on experimental setups $a$ and $b$ of both subsystems and also on $\psi$ \cite{context}. The explicit dependence of the local averages on experimental settings of the other wing is washed away after integrating out $\lambda$ so the non-signaling conditions are satisfied. $\langle A(a)\rangle_\psi$ is the average given by quantum mechanics. The distribution of the underlying ontic states is written as $\rho_\psi(a,b,\lambda)$ to account for possible retrocausal effects by measurements on the physical state of the system. For the correlation between $A(a)$ and $B(b)$\cite{comment}
\begin{eqnarray}
\int A_{\psi}(a,b,\lambda)B_{\psi}(a,b,\lambda)\rho_{\psi}(a,b,\lambda)d\lambda=\langle A(a)B(b)\rangle_\psi, \label{AB_ave}
\end{eqnarray}
where $\langle A(a)B(b)\rangle_\psi$ is again given by quantum mechanics. \\

Suppose $\lambda$ can be decomposed into a pair of variables $(\mu,\tau)$ and write $\rho_\psi(a,b,\lambda)=\rho_{\psi,\tau}(a,b,\mu)\rho_{\psi}(a,b,\tau)$, where $\mu$ and $\tau$ represent inaccessible and accessible part of the ontic state respectively. Suppose after integrating out $\mu$, we obtain local averages of $A$ and $B$ of the effective ontological model with accessible ontic information $\tau$ as follows
\begin{eqnarray}
\int A_{\psi}(a,b,\lambda)\rho_{\psi,\tau}(a,b,\mu)d\mu &=& f_\psi(a,\tau), \nonumber \\
\int B_{\psi}(a,b,\lambda)\rho_{\psi,\tau}(a,b,\mu)d\mu &=& g_\psi(b,\tau). \label{eff_ave}
\end{eqnarray}
The effective local average of $A$ given by $f_\psi(a,\tau)$ is now independent of $b$ and similarly for $B$. Hence the non-signaling conditions are satisfied while allowing for access to $\tau$. In principle, $f_\psi(a,\tau)$ and $g_\psi(b,\tau)$ could give different outcome distributions from quantum mechanics. \\
 
Now I'll focus on a pair of qubits and study the general constraints set by quantum mechanics on the possible deviations of $f_\psi(a,\tau)$ and $g_\psi(b,\tau)$ from  quantum mechanical distributions. Following \cite{extension}, the variance of $f_\psi(a,\tau)$ over $\rho_\psi(\tau)$ is defined as
\begin{eqnarray}
\delta_\psi(a)=\int(f_\psi(a,\tau)-\langle A(a)\rangle_\psi)^2\rho_\psi(\tau)d\tau, \label{variance_def}
\end{eqnarray}
where I have omitted the possible dependence of the ontic state distribution on $a$ and $b$ for simplicity of notation. The local spin observable $A(a)$ takes two discrete values $\pm1$. Since $-1\leq f_\psi(a,\tau)\leq1$, $\delta_\psi(a)$ can be upper bounded as
\begin{eqnarray}
\delta_\psi(a)\leq\int |f_\tau(a,\tau)|\rho_\psi(\tau)d\tau-\langle A(a)\rangle^2_\psi. \label{variance_bound1}
\end{eqnarray}
Consider a series of local spin measurements in directions given by $2n+1$ unit vectors $\gamma_j$, $j=0,\cdots,2n$ with $\gamma_0=a=-\gamma_{2n}$. Perform the following local measurements with the local observables $A(\gamma_j)$, $B(\gamma_{j+1})$ when $j$ is even; and  $A(\gamma_{j+1})$, $B(\gamma_j)$ when $j$ is odd. Since $A(\gamma_j)=\pm1$ and $B(\gamma_j)=\pm1$, for a general ontological model, $-1\leq A_\psi(\gamma_j,\gamma_{j+1},\lambda)\leq1$ where equalities hold when the model is deterministic. We have the following inequality
\begin{eqnarray}
|A_\psi(\gamma_j,\gamma_{j+1},\lambda)-B_\psi(\gamma_j,\gamma_{j+1},\lambda)|\leq1-A_\psi(\gamma_j,\gamma_{j+1},\lambda)B_\psi(\gamma_j,\gamma_{j+1},\lambda), \label{ineq_A-B}
\end{eqnarray}
from which the inequality follows
\begin{eqnarray}
|f_\psi(\gamma_j,\tau)-g_\psi(\gamma_{j+1},\tau)|\leq1-E_{\psi,\tau}(\gamma_j,\gamma_{j+1}), \label{ineq_f-g}
\end{eqnarray}
where
\begin{eqnarray}
E_{\psi,\tau}(a,b)\equiv \int A_\psi(a,b,\lambda)B_\psi(a,b,\lambda)\rho_{\psi,\tau}(a,b,\mu)d\mu. \label{def_E}
\end{eqnarray}
The rest of the derivation is the same as in \cite{extension} and the same constraint on $\delta_\psi(a)$ is obtained 
\begin{eqnarray}
\delta_\psi(a)\leq \mathrm{min}_{\gamma_1,\cdots,\gamma_n}\Omega_\psi(a,n)-\langle A(a)\rangle^2_\psi, \label{variance_bound2}
\end{eqnarray}
where 
\begin{eqnarray}
\Omega_\psi(a,n)\equiv n-\frac{1}{2}\sum_{k=0}^{n-1}(\langle A(\gamma_{2k})B(\gamma_{2k+1})\rangle_\psi+\langle A(\gamma_{2k+2})B(\gamma_{2k+1})
\rangle_\psi).\label{def_Omega}
\end{eqnarray}
When $\psi$ is a maximally entangled state, both terms in the r.h.s. of eqn. (\ref{variance_bound2}) vanish, and $f_\psi(a,\tau)=\langle A(a)\rangle_\psi$ for all $a$. For a generally entangled state, quantum mechanics sets an upper bound on the possible deviation of $f_\psi(a,\tau)$ from $\langle A(a)\rangle_{\psi}$ in the form of the mean squared variance $\delta_\psi(a)$. The upper bound is in general larger than zero and will be studied in detail in the next section. \\

In the above derivation, no assumption about free measurement choice is needed and the only assumption needed is the non-signaling conditions satisfied by local observers with access to partial information $\tau$ of the underlying ontic state $\lambda$. In addition, any general ontological model, deterministic or not, including the possibility of retrocausality, satisfies the same constraint (\ref{variance_bound2}) set by quantum mechanics obtained in \cite{extension}. \\

Hence, given the non-signaling conditions, no alternative theory can have improved predicability for local measurements on a maximally entangled qubits. On the other hand, when the local measurement outcomes are non-uniformly distributed (corresponding to a non-maximally entangled qubits\cite{GR_general}), it is possible for an alternative theory to yield improved predicability which is constrained by quantum mechanics. It's worth noting that the extensibility of quantum mechanics is unaffected by imposing additional assumptions concerning free measurement choices (e.g., FR and FW assumptions). \\

\section{How much better can an alternative theory do than quantum mechanics?} \label{improvement}

In light of the possibility for an alternative theory to do better than quantum mechanics in most situations, a natural question arises: given the constraint (\ref{variance_bound2}) set by quantum mechanics, what is the maximal predicability achievable by an alternative theory and what is the corresponding measurement outcome statistical distribution? In this section, I will try to address this question from two aspects. I will first study the the optimal statistical distribution given a fixed $\delta_\psi(a)$. I will then study the maximal $\delta_\psi(a)$ for a class of ontological models. \\

\subsection{Optimal statistical distribution} \label{optimal distribution}
Since $\delta_\psi(a)$ gives the mean square variance of a local measurement outcome distribution, an obvious question to raise is how the mean square variance is related to the predicability on the measurement outcomes. A good candidate of measure for the predicability is the information entropy associated to measurement outcome distributions. If the measurement outcomes are certain, the corresponding theory gives deterministic prediction of the measurement outcomes and the entropy vanishes. On the other end, if the measurement outcomes are uniformly distributed, then the theory is unable to give any prediction of the measurement results and the entropy is maximal. \\

Here I consider the family of the R$\mathrm{\acute{e}}$nyi entropy $H_{\alpha}(X)=\frac{1}{1-\alpha}\mathrm{log}(P^{\alpha}_a(X=1)+P^{\alpha}_a(X=-1)), 0\leq\alpha\leq \infty$ associated to the local measurement $A(a)$ outcomes $P_a(X=1)$ and $P_a(X=-1)=1-P_a(X=1)$, where $P_a(X=\pm1)$ corresponds to probability of projecting the local spin to direction $\pm a$ respectively. For a distribution of $f_{\psi}(a,\tau)$ over $\rho_{\psi}(\tau)$, the averaged entropy is 
\begin{eqnarray}
\bar{H}_{\alpha}&=&\int H_{\alpha}(X(\tau))\rho_{\psi}(a,\tau) d\tau \nonumber \\
&=& \int \frac{1}{1-\alpha}\mathrm{log}((\frac{1+f_{\psi}(a,\tau)}{2})^{\alpha}+(\frac{1-f_{\psi}(a,\tau)}{2})^{\alpha})\rho_{\psi}(a,\tau)d\tau. \label{H_ave}
\end{eqnarray} 
We want to find the minimal $\bar{H}_{\alpha}$ given the variance $\delta_{\psi}(a)$ of $f_{\psi}(a,\tau)$. Note that the average value of $f_{\psi}(a,\tau)$ is $\langle A(a)\rangle_{\psi}$. Furthermore, $f_{\psi}(a,\tau)$ is supposed to provide more information than $\langle A(a)\rangle_{\psi}$ since the later is coarse-grained from the former. Hence, the entropic function should be concave so that $\bar{H}_\alpha \leq H_\alpha(X_{\psi})$ where $X_{\psi}$ referrs to the probability distribution given by quantum mechanics. In light of the concavity requirement, the min-entropy $H_{\infty}(X)$ is not suitable for characterizing the predicability of an alternative theory on measurement outcomes. All other members of the R$\mathrm{\acute{e}}$nyi entropy are valid candidates. \\

It turns out that in general, the minimal $\bar{H}_{\alpha}$ and the corresponding distribution $f_{\psi}(a,\tau)$ over $\rho_{\psi}(\tau)$ depend on both the specific member of $H_{\alpha}(X)$ and $\delta_{\psi}(a)$. Nevertheless, for $\alpha\leq1$ and $\delta_{\psi}(a)\leq \delta_{\psi,c}(a)$, the minimal averaged entropy is always achieved when $P_a(X=1)$ takes two values over $\tau$: 
\begin{eqnarray}
P_{a,1}(X=1)&=&0, \nonumber \\
P_{a,2}(X=1)&>&\frac{1+\langle A(a)\rangle_{\psi}}{2}, \label{bi_dist<}
\end{eqnarray}
when $\langle A(a)\rangle_{\psi}<0$; 
\begin{eqnarray}
P_{a,1}(X=1)&=&1, \nonumber \\
P_{a,2}(X=1)&<&\frac{1+\langle A(a)\rangle_{\psi}}{2}, \label{bi_dist>}
\end{eqnarray}
when $\langle A(a)\rangle_{\psi}>0$. In these situations\cite{comment_limit}, the optimal performance among all theories in predicting the measurement outcomes is given by a bi-local distribution of $P_a(X=1)$: some $\tau$ predict a deterministic outcome and other $\tau$ a probabilistic outcome. \\

The critical value of the variance $\delta_{\psi,c}(a)$ monotonically decreases with $\alpha$ and is lower bounded as \cite{comment_limit2}: when $\langle A(a)\rangle_{\psi}<0$,
\begin{eqnarray}
\delta_{\psi,c}(a)\geq 4P_{\psi,a}(X=1)(\frac{1}{2}-P_{\psi,a}(X=1)), \label{delta_critical}
\end{eqnarray}
where $P_{\psi,a}(X=1)$ is the probability given by quantum mechanics; when $\langle A(a)\rangle_{\psi}>0$, the r.h.s of (\ref{delta_critical}) takes a symmetrical form
with $X=1$ replaced by $X=-1$. As $\alpha$ approaches zero, $\delta_{\psi,c}(a)$ approaches its maximal possible value corresponding to fully deterministic measurement outcomes: $P_{a,1}(X=1)=0$ and $P_{a,2}(X=1)=1$. Thus, for the R$\mathrm{\acute{e}}$nyi entropy towards max-entropy $H_0(X)$, the constraint on $\delta_{\psi}(a)$ is disappearing and the above bi-local distribution (\ref{bi_dist<}) and (\ref{bi_dist>}) of outcome probabilities yields the minimal entropy for essentially all given $\delta_{\psi}(a)$. \\


\subsection{Maximal variance} \label{maximal variance}

I have considered the mathematical problem of minimizing information entropy among theories with the same deviation $\delta_{\psi}(a)$ from quantum mechanics. Now, I will study the maximal $\delta_{\psi}(a)$ by considering a class of ontological models. \\

The models originate from the ones constructed in \cite{extension}. Consider a generally entangled state $|\psi\rangle=\mathrm{sin}\frac{\theta}{2}|00\rangle+\mathrm{cos}\frac{\theta}{2}|11\rangle$ with $0\leq\theta\leq\frac{\pi}{2}$. The ontic state is identified with $\{\psi,\lambda\}$ \cite{comment_ontic}, where $\lambda$ is a unit vector in the 3-dimensional real space. The local observables are given by $\hat{A}(a)=a\cdot\sigma$ and $\hat{B}(b)=b\cdot\sigma$, where $a$ and $b$ are unit vectors representing directions of local spin measurements and 
$\sigma=(\sigma_x,\sigma_y,\sigma_z)$ is the vector of Pauli matrices. $\sigma_z$ yields eigenstates $|1\rangle$ and $|0\rangle$ with eigenvalues $\pm1$. Assume $a$ and $b$ lie in the plane orthogonal to the direction of propagation of the entangled pair. The response of $\{\psi,\lambda\}$ to measurement $A(a)$ and $B(b)$ are given by
\begin{eqnarray}
A_{\psi}(a,b,\lambda)=
\begin{cases}
+1, &\text{if $\tilde{a}\cdot\lambda\geq\mathrm{cos}\hspace{1pt}\xi$}, \\
-1, &\text{if $\tilde{a}\cdot\lambda<\mathrm{cos} \hspace{1pt}\xi$},
\end{cases} \label{A_response}
\end{eqnarray}
and 
\begin{eqnarray}
B_{\psi}(a,b,\lambda)=
\begin{cases}
+1, &\text{if $b\cdot\lambda\geq\mathrm{cos} \hspace{1pt}\chi$}, \\
-1, &\text{if $b\cdot\lambda<\mathrm{cos} \hspace{1pt}\chi$},
\end{cases} \label{B_response}
\end{eqnarray}
where $\mathrm{cos}\hspace{1pt} \xi=-\langle A(a)\rangle_{\psi}$, $\mathrm{cos} \hspace{1pt}\chi=-\langle B(b)\rangle_{\psi}$, and $\tilde{a}=\tilde{a}(a,b)$ lies in the plane of $a$ and $b$. $\tilde{a}$ depends on both $a$ and $b$ in such a way that $A_{\psi}(a,b,\lambda)B_{\psi}(a,b,\lambda)$ averaged over $\rho_{\psi}(\lambda)$ gives the correct quantum mechanical correlation $\langle A(a)B(b)\rangle_{\psi}$. The model reproduces quantum mechanical predictions when averaged over $\rho_\psi(\lambda)=1/4\pi$. \\

Since $\lambda$ corresponds to a unit vector on a unit sphere, it can be written in spherical coordinates $(\mu,\tau)$ with pole identified by the direction of the incoming qubit from the pair. When integrating out the azimuthal angle $\mu$, the inaccessible part of $\lambda$, the resulting local average is non-signaling. When $\langle A(a)\rangle_{\psi}<0$, $-1<f_{\psi}(a,\tau)<0$ if $|\tau-\frac{\pi}{2}|<\xi$ and $f_{\psi}(a,\tau)=-1$ otherwise (and similarly for $\langle A(a)\rangle_{\psi}>0$). A similar expression holds for $g_{\psi}(b,\tau)$. We clearly see $f_{\psi}(a,\tau)$ (and $g_{\psi}(b,\tau)$) yield prediction different from quantum mechanics. Without loss of generality, I'll exclusively discuss local measurement outcomes for $A(a)$ and $\langle A(a)\rangle_{\psi}<0$ unless stated otherwise. \\

As noted by GR,  the upper bound by eqn. (\ref{variance_bound1}) is not saturated by $f_{\psi}(a,\tau)$. To saturate it, $f_{\psi}(a,\tau)$ should take values in the set $\{-1,0,1\}$. This can be readily achieved by rearranging the supports of $A_{\psi}(a,b,\lambda)$ and $B_{\psi}(a,b,\lambda)$ on the unit sphere. Let $A_{\psi}(a,b,\lambda)=-\mathrm{sign}\langle A(a)\rangle_{\psi}$ for $\mu_{\tilde{a}'}-\frac{\pi}{2}<\mu<\mu_{\tilde{a}'}+\frac{\pi}{2}$ ($\mu$ span an angle of $
\pi$ centered around $\mu_{\tilde{a}'}$) and $|\tau-\frac{\pi}{2}|<\eta$, and $ A_{\psi}(a,b,\lambda)=\mathrm{sign}\langle A(a)\rangle_{\psi}$ otherwise, where $\eta$ is a function of $\langle A(a)\rangle_{\psi}$, $\mu_{\tilde{a}'}$ is the azimuthal angle of $\tilde{a}'=\tilde{a}'(a,b)$ which plays the same role of $\tilde{a}$ previously defined. A similar expression can be found for $B_{\psi}(a,b,\lambda)$. With the rearrangement, $f_{\psi}(a,\tau)=0$ for $|\tau-\frac{\pi}{2}|<\eta$ and $f_{\psi}(a,\tau)=\mathrm{sign}\langle A(a)\rangle_{\psi}$ otherwise. The modified model saturates the bound of (\ref{variance_bound1}), leading to the variance
\begin{eqnarray}
\delta_{\psi}(a)=|\langle A(a)\rangle_{\psi}|-\langle A(a)\rangle^2_{\psi}. \label{variance_bound3}
\end{eqnarray}

GR noted that the modified model saturates the bound (\ref{variance_bound2}) set by quantum mechanics for $\sigma_z$. But the strict bound for spin measurement other than $\sigma_z$ is left unaddressed. As the class of ontological models considered here predict outcomes of spin measurements in all possible directions on an equal footing, it is natural to wonder whether (\ref{variance_bound3}) is the strict bound for all $a$. It is indeed the case for a class of ontological models which rearrange the supports of $A_{\psi}(a,b,\lambda)$ and $B_{\psi}(a,b,\lambda)$ on the unit sphere in all possible ways. \\

The original model with supports given by eqn. (\ref{A_response}) and (\ref{B_response}) yields a distribution of $P_{a,\tau}(X=1)<\frac{1}{2}$ which is a continuous function of $\tau$ for $|\tau-\frac{\pi}{2}|<\xi$ and is deterministic $P_a(X=1)=0$ otherwise. The corresponding variance $\delta_{\psi}(a)$ is smaller than the above modified supports associated with a bi-local distribution $P_{a,1}(X=1)=0$ and $P_{a,2}(X=1)=\frac{1}{2}$. The supports given by eqn. (\ref{A_response}) and (\ref{B_response}) correspond to circularly bounded regions on the unit sphere. The modified supports (associated with $X=1$) distort the shape of circular areas to belt-shaped regions wrapping around half of the equator. After integrating out the azimuthal angle, the two different shapes of supports give rise to different distributions of $P_a(X=1)$. \\

Geometric considerations suggest that supports with belt-shape are optimal for maximizing the distribution variance. Within the class of ontological models discussed here, the only way to surpass the bound of (\ref{variance_bound3}) is to compress the belt-shaped regions so that they span more than half of the equator yielding a distribution $P_{a,1}(X=1)=0$ and $P_{a,2}(X=1)>\frac{1}{2}$ (or more generally, $P_a(X=1)=0$ for some $\tau$ and $P_{a,\tau}(X=1)>1/2$ for some other $\tau$). However this is impossible as can be readily proved. Without loss of generality, let's consider the bi-local distribution (the cases of continuous distributions follow the same reasoning). There are two scenarios: 1) both of $P_{a,2}(X=1)$ and $P_{b,2}(X=1)$ are larger than $1/2$; 2) only one of them, say $P_{a,2}(X=1)$ is larger than $1/2$. \\

In case 1), the conditional probability $P(X_b=1|X_a=1)>0$ for all $a$ and $b$ due to non-vanishing overlap between supports of $A_{\psi}(a,b,\lambda)=1$ and $B_{\psi}(a,b,\lambda)=1$ around the equator. On the other hand, quantum mechanics predicts the existence of vanishing conditional probability for any $a$ by choosing $b$ properly. Hence case 1) is inconsistent with quantum mechanics and is ruled out. Similarly in case 2),  $P(X_b=1|X_a=1)<1$ for all $a$ and $b$ since the two belt-shaped regions span different azimuthal angles. While quantum mechanics predicts $P(X_b=1|X_a=1)=1$ for any $a$ by orienting $b$ according to $a$. Thus case 2) is also ruled out. So the bound (\ref{variance_bound3}) is the strict bound for all $a$ within the class of ontological models derived from the models in \cite{extension}. In particular, the proof also applies to retrocausal version of the discussed models for which the distribution of $\lambda$ can be affected by both $A(a)$ and $B(b)$. \\



\section{Conclusions and discussion} \label{Conclusions}

In this paper, I clarify the dispute between CR and GR, and point out a mistake in the proof of the CR no-go theorem. I then show that the extensibility of quantum mechanics is insensitive to assumptions concerning the freedom of measurement choices. In fact, the constraint (\ref{variance_bound2}) set by quantum mechanics can be derived assuming only the non-signaling conditions for local observers with partial access to the underlying ontic states. \\

I further study the optimal predicability on the local spin measurement outcomes of a pair of entangled qubits. For the variance $\delta_{\psi}(a)<\delta_{\psi,c}(a)$ of the local measurement outcomes, I show that the minimal entropy of measurement outcomes can be achieved by bi-local distributions (\ref{bi_dist<}) and (\ref{bi_dist>}) for the R$\mathrm{\acute{e}}$nyi entropy $H_{\alpha}(X)$ of $\alpha\leq1$. I prove for a class of ontological models, (\ref{variance_bound3}) is the strict upper bound on the derivation from quantum mechanics for a general local spin measurement. \\

Although the ontological models in section \ref{maximal variance} are defined on a unit sphere, the resulting outcome statistics is more general, independent of the specific spherical geometry. Hence, I suspect that the proof of the upper bound is more generally valid and conjecture (\ref{variance_bound3}) to be the strict bound for all local spin measurements among all ontological models. \\

Finally, it's worth pointing out that in the discussion of section \ref{maximal variance}, $\sigma_z$ is of special significance physically than other local measurements since it projects onto the diagonal basis of the entangled state $|\psi\rangle=\mathrm{sin}\frac{\theta}{2}|00\rangle+\mathrm{cos}\frac{\theta}{2}|11\rangle$. This is the starting point of the analysis of the extensibility of quantum mechanics on predicting local measurement outcomes. So the strict constraint set by quantum mechanics on predicting a general local 2-outcome measurement is given by (\ref{variance_bound3}) already proved in \cite{extension}. \\

\section*{Acknowledgement}

The author thanks the support from the Department of Computer Science \& Engineering, Texas A\&M University.


\begin{thebibliography}{99}
\bibitem{consistent} An additional (and natural) assumption needed is that any extra measurement $C$ with outcome $Z$ is consistent with the prediction of quantum mechanics, namely, $\sum_zP_{z|C}P_{XY|ABCz}=P_{XY|AB}$.
\bibitem{comment} I will focus on of a pair of qubits, so it suffices to consider local averages $A$ and $B$ and their joint average. Consistency with quantum mechanics for these averages guarantees the same outcome statistics between an ontological model after integrating out ontic states and quantum mechanics.
\bibitem{comment_limit} The actual range of $\alpha$ for which (\ref{bi_dist<}) and (\ref{bi_dist>}) apply is larger, with the critical value $\alpha_c$ around 1.38. I adopt $\alpha\leq1$ in the text since $\alpha=1$ corresponds to the most commonly used Shannon entropy.
\bibitem{comment_limit2} The actual range of $\delta_{\psi}(a)$ for which (\ref{bi_dist<}) and (\ref{bi_dist>}) apply is larger, with the lower bound larger than (\ref{delta_critical}): $\delta_{\psi,c}(a)= 4P_{\psi,a}(X=1)(\frac{1}{2}+\eta-P_{\psi,a}(X=1))$, $\eta>0$. In light of the discussion on the maximal variance in section \ref{maximal variance}, it suffices to adopt (\ref{delta_critical}) since it already saturates the maximal variance.
\bibitem{comment_ontic} Previously, I have identified $\lambda$ with ontic state for generality. Here for convenience, I adopt GR's definition \cite{extension} and take the pair $\{\psi,\lambda \}$ as the ontic state.
\end{thebibliography}
\end{document}